\documentclass[conference]{IEEEtran}
\IEEEoverridecommandlockouts

\usepackage{amsmath,amssymb,amsfonts}
\usepackage{algorithmic}
\usepackage{graphicx}
\usepackage{textcomp}
\usepackage{xcolor}
\usepackage[utf8]{inputenc}
\usepackage{comment}

\usepackage[
backend=biber,
style=numeric,
sorting=nty
]{biblatex}
\begin{document}

\title{Differential Privacy Made Easy\\
}

\author{\IEEEauthorblockN{Muhammad Aitsam}
\IEEEauthorblockA{ \\
\textit{Sheffield Hallam University}\\
\textit{United Kingdom} \\
m.aitsam@shu.ac.uk}

}
\maketitle

\begin{abstract}

Data privacy is a major issue for many decades,
several techniques have been developed to make sure individual’s privacy but still world has seen privacy failures. In 2006, Cynthia Dwork gave the idea of Differential Privacy which gave strong theoretical guarantees for data privacy. Many companies and
research institutes developed differential privacy libraries, but in order to get the differentially private results, users have to tune the privacy parameters. In this paper, we minimized these tune-able parameters. The DP-framework is developed which compare the differentially private results of three Python based DP libraries. We also introduced a new very simple DP library $(GRAM-DP)$, so the people with no background of differential privacy can still secure the privacy of the individuals in the dataset while releasing statistical results in public.
\end{abstract}


\section{Introduction}

For many years, personal data is being collected by the government, hospitals, companies for different purposes, which sometimes help them in decision making to set or achieve their organizational goals. However, this data might be shared or sold to other organizations for further analysis. This data sharing helps researchers and companies to analyze data again for potential applications, but due to the presence of sensitive information in the dataset, it is crucial to preserve the privacy of an individual. The problem is how to analyze and share statistical data without compromising an individual's privacy. In a setting where the curator owns a database containing specific information, a privacy breach occurs when an adversary infers this information. An adversary can use different techniques or preys on the background knowledge, even when the released data is anonymized.

While querying the database, there is a possibility that data analysts get knowledge about someone whose data is present in the database. This leakage of user privacy is undesirable. That is why there is a need for a system that can protect an individual's privacy while sharing useful statistics. Many techniques are developed to achieve this goal, but we have seen privacy protection failures that lead to the re-identification of individuals e.g., Insurance Commission (GIS), and Netflix Award. Some privacy-preserving mechanisms like Data Anonymization \cite{anony} and Data Encryption \cite{li} were proposed earlier, but later on, research shows that these methods could not stand some well-known attacks like Homogeneity Attack \cite{homo}, Background Knowledge Attack \cite{knowledge}, and Inference Attack \cite{infer}.

Our contributions in this paper are as follows:
\begin{itemize}
    \item Summarizing all the basic points of Differential Privacy.
    \item Simplified DP-framework where user can compare the results of three python-based Differential Privacy Libraries.
    \item GRAM-DP: simple differential privacy library with minimum parameter requirement. 
\end{itemize} 

This paper provides a basic understanding to the people who are new in the field of Differential Privacy $(DP)$. In the next section, we discuss the privacy failures in past to understand the importance of Differential Privacy. In Section 3 and 4, we introduce Differential Privacy along with its formal definitions. Section 5 is about some important differential privacy mechanisms. In Section 6, we present our DP-framework. In Section 7, we present GRAM-DP and its usage. Section 8 is about use cases and discussions on results we got.

\section{Privacy Failures in Past}

Every day a tremendous amount of data is collected and shared in multiple ways, which led to an increase in privacy concerns. Many studies in the recent past showed that most of the US population could be uniquely distinguished by joining their date of birth, gender, and zip code sweeny, golli]. If the individual is identified in the released data, their private data will be disclosed. The disclosure of personal information can be harmful to the user and it is also against GDPR.
It is tough to protect the dataset from background knowledge attack. It is impossible to predict what background information the adversary has, which makes privacy preservation models vulnerable. This section will discuss some famous privacy failures, which started a big debate regarding these issues.

\subsection{AOL Privacy Debacle}
In 2006, AOL disclosed anonymized data of over 650,000 users. The disclosure was done without users' consent because the AOL team thought they anonymized data well enough not to threaten anybody's Privacy. But, the New York Times demonstrated \cite{aol} that individuals can be identified based on these search queries.
After this story was released, AOL removed that data from the site and apologized for its release. But copies of these records continue to circulate online, risking the Privacy of many Americans.

\subsection{Insurance Commission (GIS)}
The MIT graduate student Latanya Sweeny has shown that an individual's privacy can be compromised even when the released data is anonymized. She demonstrated this by revealing the medical records of Massachusetts' Governor. She accomplished this task by using some background knowledge and then linking the anonymized data released by the hospital with the publicly available data (e.g., voter registration list).

\subsection{Netflix Prize}
In 2006, one of the most extensive online DVD rental services, NETFLIX, released an anonymized dataset containing 100 million movie ratings provided by 500,000 of its subscribers. They started a competition called NETFLIX Prize to develop a better movie recommendation system.
NETFLIX claimed that they secured Privacy by removing all the personal information and only kept information like unique user ID, ratings and dates the subscriber rated the movie. This time students from the University of Texas, Narayanan and Shmatikov, demonstrated that with very little background knowledge about the subscriber and linking the data with publicly available data, an individual's privacy could be compromised. They used IMDB (The Internet Movie Database) as a source of background knowledge. By linking this database with the NETFLIX database, they were able to re-identify many subscribers. After this research was published, NETFLIX canceled its second competition \cite{netflix}.

S.Ezzini et al. \cite{wheel} showed that it is possible to re-identify the users from her vehicle sensor data in the automotive industry. Another research by S.Lastyan et al. \cite{can} showed that extracting vehicle sensor signals from can logs can lead to users’ re-identification.

All mentioned privacy failures are proof that even when the released data is anonymized, there is still the possibility of a privacy breach. This privacy breach can be harmful to the users participating in the study. To motivate maximum users for participating in any particular study, there was a need for better techniques to give the participants confidence that their privacy will not be compromised whatsoever. All the above-mentioned privacy attacks can be neutralized and all these privacy failures could be prevented with Differential Privacy.

\section{Differential Privacy}

The first formal definition of statistical data privacy covering all essential aspects was given by Dalenius in 1977 \cite{cynthia}. According to him, anything that can be learned from the statistics about an individual should be determined with or without access to the database \cite{cynthia}. In other words, the difference between learning something new when someone is in the dataset and when someone is not in the dataset should be very tiny. This is precisely a definition of semantic security \cite{semantic}. Practically, it is impossible to achieve this tiny difference because here, the advisory and legitimate recipient is the same person (data analyst). He/ She might have good reasons to get the results, or they could also have bad motivations. To solve this issue and to limit the harm to the teachings of the database instead of somebodies participation in it, Cyntia Dwork came up with a new definition of Privacy: the risk of an individual's privacy (e.g., risk of being denied automobile insurance) should not substantially increase due to participating in a database \cite{promise}. The method through which the privacy preserved statistical analysis is released is known as Differential Privacy.

Differential Privacy gives a firm definition for data privacy. It is based on the idea that the outcome of the statistical analysis is equally likely independent of whether an individual joins or refrains from joining the dataset. This statement covers all the requirements for data privacy. An adversary can bring harm or good to any individual or group regardless of their presence in the dataset. Due to this property of Differential Privacy, it is considered a promising privacy-preserving method. Since the birth of Differential Privacy, researchers have explored many corner cases in the theoretical world. For practical cases, there is still a vast area to explore. Usually, it is done by adding noise to the query results. This section will discuss the importance of Differential Privacy, its formal definition, properties, mechanisms, and some common statistical operations.

Differential Privacy encourages users to share their data for statistical analysis by promising them that the adversary will not be able to re-identify them irrespective of the auxiliary information he/she has. Differential Privacy claims that nothing could be learned about an individual while learning useful information about a population.

Compared to other privacy-preserving models we discussed in the previous section, the Differential Privacy definition covers all aspects of data privacy and provides strong theoretical guarantees for statistical analysis. 
With the passage of time, The General Data Protection Regulations (GDPR) are getting tough because sensitive data is becoming more vulnerable. That is why modern research on statistics is focusing on Differential Privacy.

\section{Definitions of Differential Privacy}

In this section, we will discuss the formal definitions of Differential Privacy. But before that, we must understand the critical parameters used in Differential Privacy.

\begin{itemize}
	\item $\varepsilon$ - The privacy parameter which can be controlled by the data analyst to maintain the trade-off between privacy and accuracy. $\varepsilon$-differential privacy is known as \emph{pure differential privacy}.
	\item $\delta$ - The parameter which tells the probability of privacy leak ($\varepsilon$, $\delta$)-differential privacy is known as \emph{approximate differential privacy}.
	\item D1 and D2 - Neighboring dataset (differ by only one element).
	
\end{itemize}

Let's have a look at two formal definitions of Differential Privacy. The first definition is $\varepsilon$-Differential Privacy also known as pure Differential privacy. Here the $\delta$ is considered to be zero and $\varepsilon$ is the only privacy parameter. The second definition is ($\varepsilon$, $\delta$)-Differential Privacy also known as approximate Differential Privacy. Here the $\delta$ value is not equal to zero and it tells the probability of privacy breach. So, for approximate Differential Privacy, we have two ($\varepsilon, \delta$) privacy parameters.

\subsection{$\varepsilon$ - Differential Privacy}

Let $\varepsilon >0$. \emph{Define a randomized function M to be ($\varepsilon$)-differentially private if for all neighboring input datasets D1 and D2 differing on at most one element and $\forall$S $\subseteq$ Range(M), we have}\cite{cynthia}.

$$\frac{Pr[M(D1)  \in  S]} {Pr[M(D2)  \in  S]} \leq e^\varepsilon$$

\emph{where the probability is taken over the coin tosses of M} \cite{dwork}. 
The above equation can also be written as:
\begin{center}
	Pr[M(D1)  $\in$  S] $\geq$ $e^\varepsilon$ .Pr[M(D2)  $\in$  S]
\end{center}

The probability of output in S on a D1 dataset is at least $e^\varepsilon$ times to the probability of output in S on a D2 datasets.

\subsection{($\varepsilon$, $\delta$) - Differential Privacy}

\emph{Define a randomized function M to be ($\varepsilon$, $\delta$)-differentially private if for all neighboring input datasets D1 and D2 differing on at most one element and $\forall$S $\subseteq$ Range(M), we have}\cite{hardt}.

\begin{center}
	Pr[M(D1)  $\in$  S] $\geq$ $exp(\varepsilon)$ $\times$ Pr[M(D2)  $\in$  S] + $\delta$
\end{center}

In the above equation, we have two privacy parameters. Epsilon ($\varepsilon$) and delta ($\delta$).
Where delta ($\delta$) is the probability of privacy leakage. 

For instance, suppose \emph{J} is an output that possibly discloses \emph{K's} identity or data, where that parallel dataset D2 does not contain \emph{K's} data, so we can say Pr[M(D2)  $\in$  S] = 0. In such case $\varepsilon$-differential privacy, \emph{M} can never output \emph{K} on any dataset, while ($\varepsilon$, $\delta$) - differential privacy may output \emph{K} with probability up to $\delta$. 

From these definitions, we can conclude that information acquired regarding the participant by the output of some algorithm should be the same or no more than the information acquired regarding the participant without accessing the output.
In a later section, we will see the practical use of these definitions.

\section{Differential Privacy Mechanisms}

Now we will discuss some famous mechanisms of Differential Privacy. They are used according to the use case. 

\subsection{Randomized Response Mechanism}

The randomized response is a case of non-interactive schemes, where each user data is perturbed individually based on the decision of coin flips This procedure provides 'plausible deniability' to the respondent \cite{random}.

\subsection{Laplace Mechanism}

Laplace mechanism is one of the most widely used mechanisms in differential privacy. In this mechanism, the random noise that adjusts to Laplace distribution with mean 0 and sensitivity GS(f)/$\varepsilon$ is added to the response of each query to make it perturbed appropriately \cite{secure}. Usually, in $\varepsilon$-differential privacy, the Laplace mechanism is used because $\varepsilon$ is the only concerned parameter when computing noise with l1 sensitivity. GRAM-DP also uses the bounded Laplace mechanism to add noise to the query answer.

\subsection{Gaussian Mechanism}

Another mechanism that has received much attention in the recent past is the Gaussian mechanism \cite{gaussian} to achieve ($\varepsilon, \delta$)-differential privacy. Here a certain amount of zero-mean Gaussian noise is added to the query result. The amount of noise added to the result scales with l2 sensitivity. Another factor that plays an important role here, along with $\varepsilon$, is $\delta$, which determines the probability of privacy leakage. Liu et al \cite{gaussian} proved a generalized Gaussian mechanism for differential privacy.

\subsection{Exponential Mechanism}

Not all queries return numerical values to their output. Hence, McSherry and Talwar \cite{design} came up with a method that can be applied to make non-numeric queries differentially private. After using the Exponential mechanism for non-numeric queries, the final output would be close to ideal answers since the mechanism appoints exponentially higher probabilities of being selected to the higher outcomes.

\section{Simplified Differential Privacy (DP) Framework}

Practical work in the field of Differential Privacy is comparatively less than theoretical work.
So far there are seven libraries (IBM-diffprivlib \cite{ibm}, OpenDP-Smartnoise \cite{smartnoise}, Openmined-PyDP \cite{pydp}, diffpriv \cite{diffpriv}, Pinq \cite{pinq}, TensorFlow-Privacy \cite{tensorNew}, Opacus-PyTorch \cite{pytorch}) which claim to confirm (global) Differential Privacy for a given dataset. In order to use the mentioned libraries, users need a proper understanding of Differential Privacy, its parameters and the working mechanism of the library. The library will return the differentially private results according to the specified parameters. For the new users, it would be hard to understand all these factors.

We developed a basic DP-framework of three differential privacy libraries (\emph{IBM-diffprivlib, OpenDP-Smartnoise, OpenMined-PyDp}). These libraries use different methods to return the DP results. In our framework we ask the user to provide only the necessary parameters and the framework will automatically arrange them according to the methods required by each library.

\subsubsection{Using DP-Framework}
As the target is make it as simple as possible for new users or people with DP backgrounds, we kept the DP-framework straightforward. To use the DP-framework one has to do the following steps:
\begin{itemize}
    \item Download the Git repository \cite{mypaper}
    \item Open desired analysis.py
    \item Specify the dataset path and the column
    \item Provide the value of $\varepsilon$
    \item Provide the value for upper bound and lower bound.  
\end{itemize}
Here $\varepsilon$ is the privacy parameter. High value of $\varepsilon$ will result in low privacy and high utility and vice versa. The user has to specify this value according to the use case. 
Bounds are also important parameters and users can guess these values according to the general information of data $\emph{e.g.}$ if the data is $'age'$ of people participating in the study then bounds could be 90 as upper-bound and 18 as lower-bound.
We also tried to minimize the parameters user has to provide. e.g. if the user has no idea about the bounds then he/she can put 'None' instead of a numeric value and DP-framework will identify them according to the maximum and minimum value in the dataset column.

\section{GRAM-DP}

The GRAM-DP, is written in Python 3, a popular programming language for data analysis. GRAM-DP leverages the functionality and familiarity of the NumPy package, meaning functions are instantly recognizable, with default parameters ensuring accessibility for all. Released under the MIT Open Source license, GRAM-DP is free to use and modify, and the contributions of its users are welcomed to help expand the functionality and features of the library.
It provides a collection of basic queries needed for statistical analysis and the fundamental building blocks of differential privacy that handle
the addition of noise.

The purpose of this library is to make differential privacy accessible for people who are investigating it for the first time. In its first release, it only contains basic queries but we are planning to add more functionalities in upcoming releases while maintaining the simplicity of the library.

\section{Basic Queries}

Let's have a look at some common statistical operations supported by GRAM-DP and their sensitivity calculation formulas that will be used to add noise to the query result. The major work to find out the equation for sensitivity calculation is done under the Harvard Open Differential Privacy project \cite{sense}.

\subsection{Count Query}

One of the most commonly used queries is the counting query. It helps to determine the total number of times a specific event happened. To make the counting query differentially private, noise is added to the outcome. As the presence or absence of any user in one particular cell can change the outcome maximum by 1, so $l1$ and $l2$ sensitivity of counting query is always 1. The Laplace mechanism amount of noise added to the correct outcome is calculated by $\frac{1}{\varepsilon}$. In counting query, for query $q$, dataset $D$ with size $n$. We use,

$$
q(D)=\frac{1}{n} \sum_{i=1}^{n} q\left(x_{i}\right)
$$
to evaluate counting queries on datasets.

\subsection{Sum Query}

It is an aggregate function that calculates the sum of desired events. It is widely used for answering statistical questions. For dataset $x$ with $n$ number of rows, Sum query will be:

$$
s(x)=\sum_{i=1}^{n} x_{i}
$$

The $l_{1}$ and $l_{2}$ sensitivity for sum query is proved in \emph{Open Differential Privacy}(openDP) white paper. According to that \emph{"Say the space of data points X is bounded above by M (maximum) and bounded below by m (minimum). Then s over $X^n$ has $l_{1}$, $l_{2}$-sensitivity in the change-one model bounded above by:"}

$$M - m$$

\subsection{Mean Query}

Another important query in data analysis is the mean query. It is an aggregate function that calculates the mean of a numerical dataset. For dataset $x$ with $n$ number of rows, Mean query will be:

$$
f(X)=\frac{1}{n} \sum_{i=1}^{n} x_{i}
$$

The maximum difference between two possible answers of the mean query is known as sensitivity of mean query. Formally it is defined as \emph{"Say the space of data points X is bounded above by M and bounded below by m. Then $f(.)$ has $l1$, $l2$-sensitivity in the change-one model bounded above by:}
\begin{center}
	$$\frac{M-m}{n}$$
\end{center}

Where $n$ is the total number of rows in dataset.

\subsection{Variance Query}

In statistics, the measurement of the spread between numbers in a dataset is known as Variance. It helps to measure how far each number in the dataset is from the mean.

$$
s^{2}(x)=\frac{1}{n-1} \sum_{i=1}^{n}\left(x_{i}-\bar{x}\right)^{2}
$$
\emph{where $\bar{x}$ refers to the sample mean of $x$}

The $l1$, $l2$-sensitivity of the Variance query is calculated by the following equation: 

\emph{For $X$ bounded between m and M, $f(.)$ has $l1$, $l2$-sensitivity in the change one model bounded above by}

\begin{center}
	$$\frac{n - 1}{n^2} (M - m)^2$$
\end{center}

\section{How GRAM-DP Works}

The goal of this library is to make use of Differential Privacy easy so absolute beginners or people from a different fields but concerned about the individual privacy of their users can get something out of it.

In order to use GRAM-DP, the user has to clone the Github repository \cite{mypaper}.
Then users can open the \emph{gram-main.py}, add the path to Comma Separated Value (csv) file and specify the column. Then all he/she has to do is to provide the level of desired privacy from \emph{very high} to \emph{very low}.

The \emph{gram-analysis.py} file looks like this:

\begin{verbatim}

#by user
array = [xxx]
query = 'xxx' 
desired_privacy = 'xxx'

def gram_dp_analysis(query,
    array, desired_privacy):
    dp_result = eval('gramdp_'+ query)
    (array=array,
    desired_privacy=desired_privacy)
    return dp_result

\end{verbatim}
User has to select one of the following privacy level.
\begin{verbatim}

desired_privacy = 'very_high'
                  'high'
                  'moderate'
                  'low'
                  'very_low'
\end{verbatim}

The following table shows the available queries and mechanism in first release of GRAM-DP, in next release we are planning to introduce more mechanisms and differential privacy features:

\begin{tabular}{ |p{3cm}||p{3cm}|}
 \hline
 \multicolumn{2}{|c|}{GRAM-DP} \\
 \hline
 \textbf{Mechanism} & \textbf{Queries}\\
 \hline
 Laplace   & Count  \\
 & Sum  \\
 & Mean \\
 & Variance \\
 \hline
\end{tabular}

\section{Use Cases}

In this section, we use our DP-framework and GRAM-DP for two real-world dataset. The selected dataset is $Adult Dataset$. It is open-source and available on the UCI machine learning repository \cite{uci}. This dataset contains enough information to compromise the privacy of the participants in the dataset. We selected two columns (age and working hours) from this dataset. See Fig.1.

\begin{figure}[!htbp]
    \centering
    \includegraphics[scale=0.2]{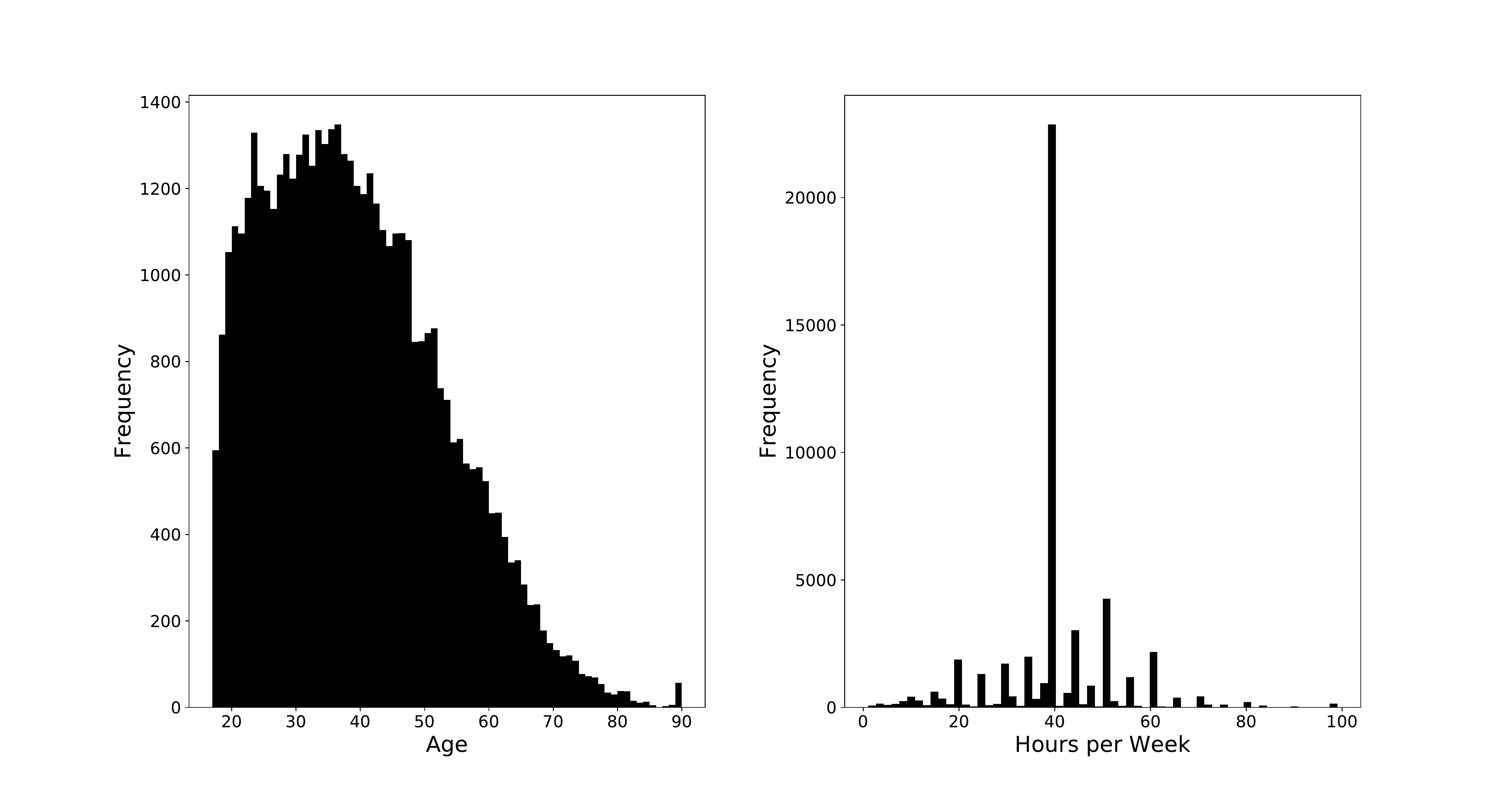}
    \caption{Frequency plot of Age and Total working hours from Adults Dataset}
\end{figure}

\subsection{DP-Framework Experiments}

To understand the differentially private results by each library we consider three types of errors, 1. Mean Scaled Error, 2. Mean Squared Error (MSE), 3. Root Mean Squared Percentage Error (RMSPE). The $\varepsilon$ value is from 0.01 to 0.5 with step size of 0.02. For each epsilon ($\varepsilon$), we ran the code for 100 iterations. Each plot in this section shows four subplots. The first one is the Average DP result, second plot shows the Mean scaled Error, third one is for MSE and last one for RMSPE.

\begin{figure}[!htbp]
    \centering
    \includegraphics[scale=0.2]{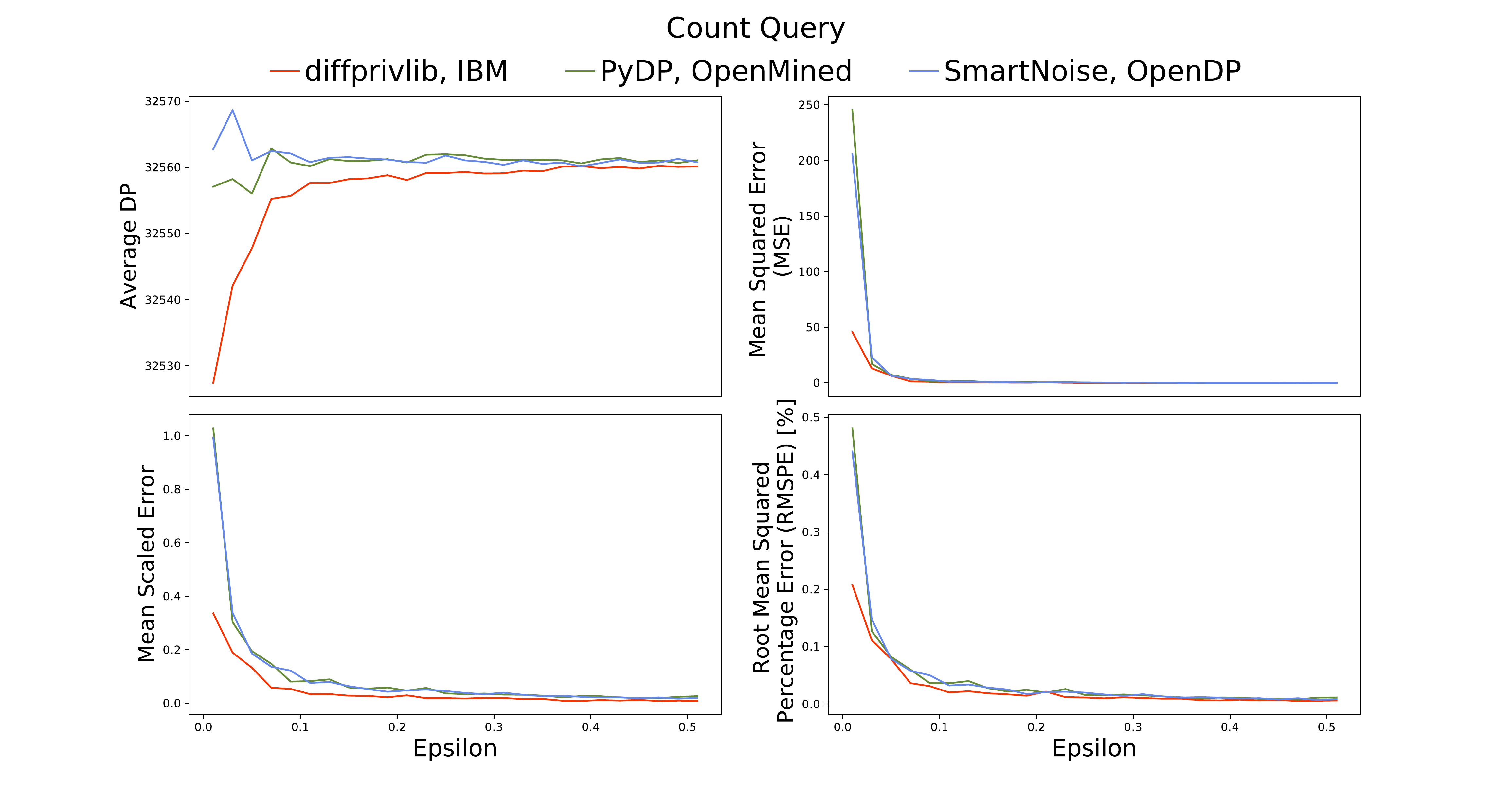}
    \caption{Count Query results for $hrs-per-week$ column in Adult dataset. The unusual behavior of diffprivlib is due to its mechanism for count query. It consider only non-zero cells in column.}
\end{figure}

\begin{figure}[!htbp]
    \centering
    \includegraphics[scale=0.2]{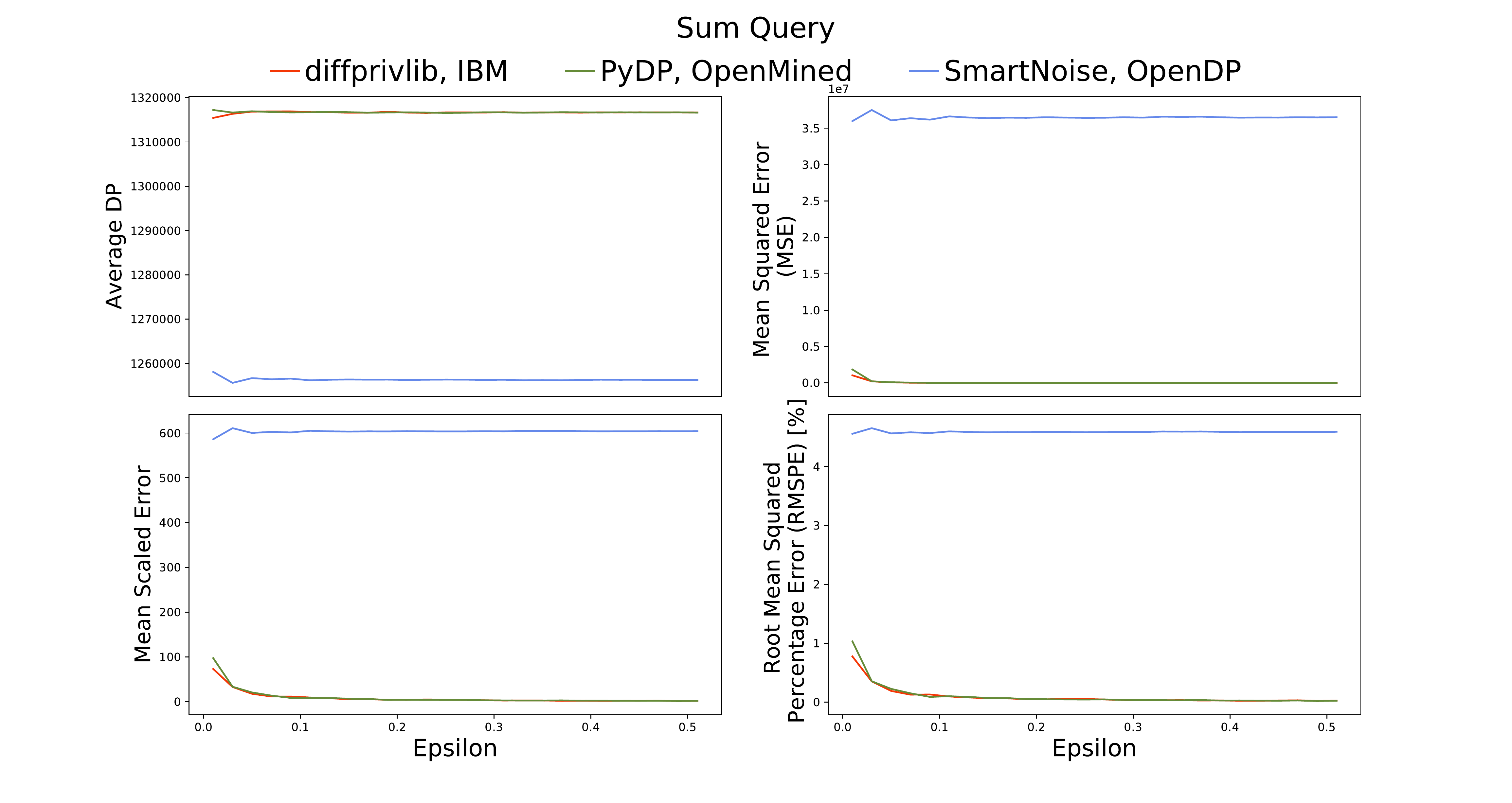}
    \caption{Sum Query results for $hrs-per-week$ column in Adult dataset. As we can see the Smartnoise performance is worst as compared to diffprivlib and PyDP.}
\end{figure}

\begin{figure}[!htbp]
    \centering
    \includegraphics[scale=0.2]{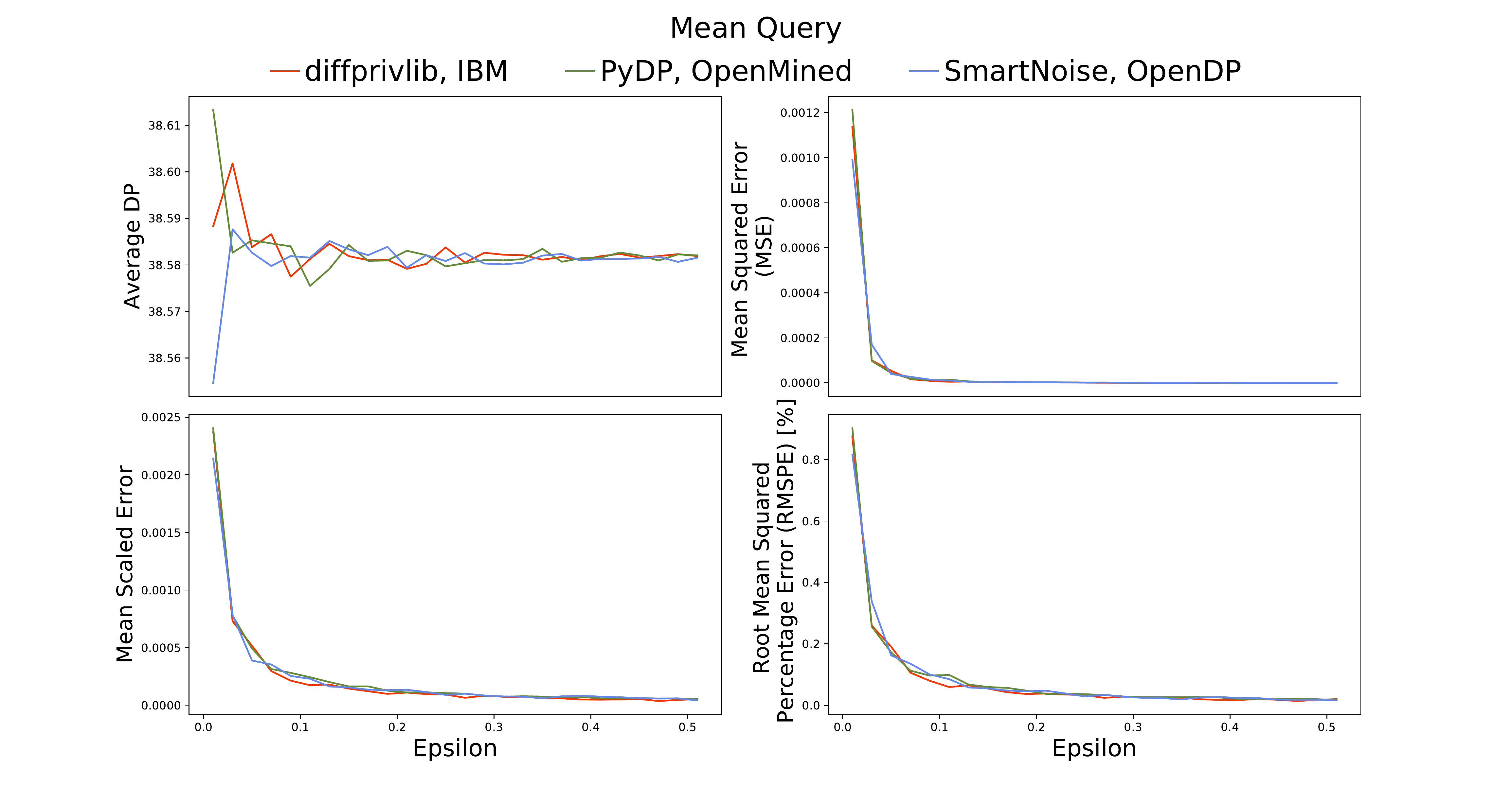}
    \caption{Mean Query results for $age$ column in Adult dataset. Results are as per our expectations for this query. With the increase in $\varepsilon$, error is decreasing for all three libraries.}
\end{figure}

\begin{figure}[!htbp]
    \centering
    \includegraphics[scale=0.2]{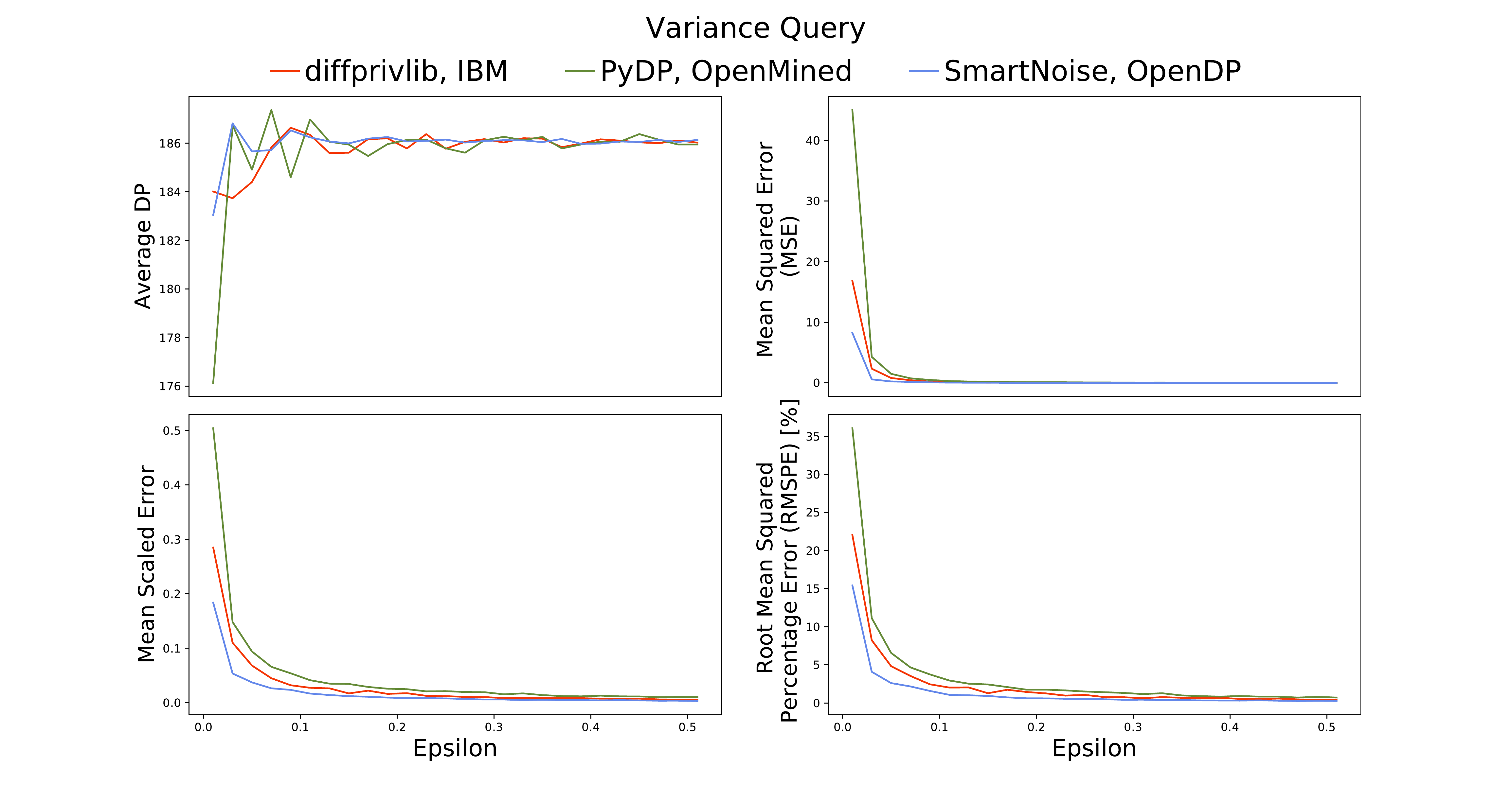}
    \caption{Variance Query results for $age$ column in Adult dataset. Mean Scaled Error, Mean Squared Error and RMSPE shows that error for PyDP is slightly higher.}
\end{figure}

 \subsection{GRAM-DP Experiments}

For $age$ column in Adult dataset we executed query (count, mean, var) for 500 iterations to 10000 iterations for every \emph{desired privacy} level. All three errors for every privacy level are plotted. The results depict how the GRAM-DP behave for every level of privacy. Here we are only discussing the interesting plots. Visit Git repository \cite{mypaper} for more plots to try out our code. 

\begin{figure}[!htbp]
    \centering
    \includegraphics[scale=0.2]{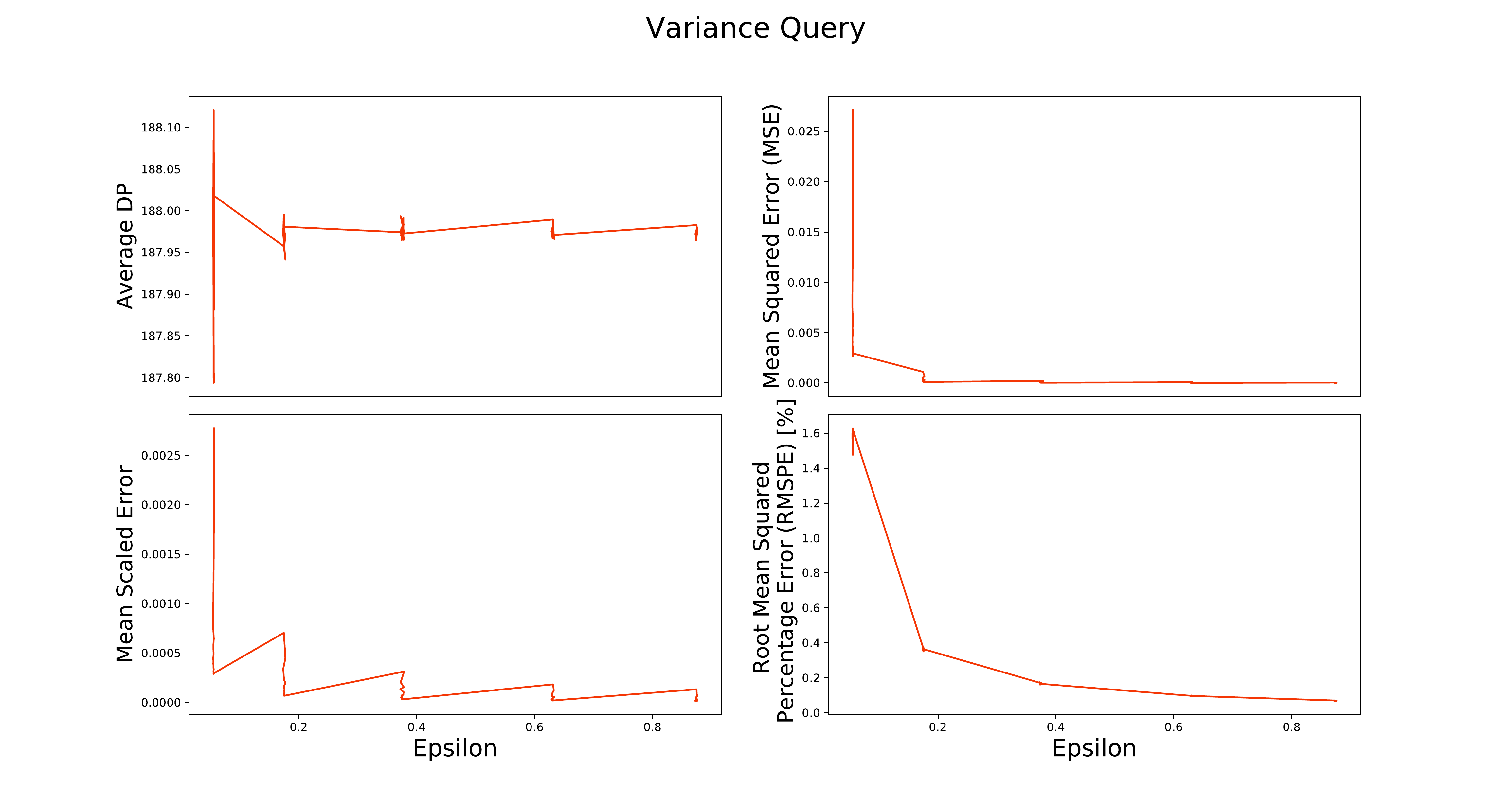}
    \caption{GRAM-DP result of Variance Query results for $age$ column in Adult dataset. Each step in all plots represents the different privacy level. Mean Scaled Error slightly increase at the end of every privacy level.}
\end{figure}

Overall results of GRAM-DP for all queries are according to our expectations. The benefit of GRAM-DP over all other Differential Privacy libraries is that it's very easy to use and users has to provide a few attributes to get the DP results.

\section{Future Work}

This paper is an effort to make Differential Privacy easy for its new users or for the people who want to use DP in their companies but don't know where to start from. A lot of work can be done to improve this project. 
For this project, we considered only the Laplace mechanism for both DP-framework and GRAM-DP. In the future more mechanisms like Gaussian, Exponential can be introduced to diversify this project. 
Another thing that can be improved is that for this project we used bounded sensitivity. One of the future tasks could be to introduce an option for unbounded sensitivity. It will help users to understand which type of sensitivity is better for their project. 
One of the future work could be to create more queries for GRAM-DP, so far we added all basic queries, but the addition of more queries will give strength to this project.
In our DP-framework we included Python based DP libraries but we are planning to include DP libraries of different programming languages in this project. We are also planning to add very specific options for each library, e.g. some libraries support 'floating point protection'. Such options will help users to tune the parameters according to the use case.

\section{Conclusion}

This paper aims to make differential privacy easy for new people in this field. We discussed all basic points related to differential privacy along with its formal definitions. For the practical implementation, we introduced DP-framework which compare three Python based differential privacy libraries and plot the outcomes for the desired value of privacy parameter ($\varepsilon$). We also published our DP library with the name GRAM-DP. GRAM-DP returns are deferentially private results for your data without asking much parameters. The experimental results for both of the contributions are according to our expectations which are shown in our Use Cases section.

\section{Git Repository}
After cloning the repository \cite{mypaper} you will see two folders. 
\begin{itemize}
    \item DP-framework
    \item GRAM-DP
\end{itemize}

Open $gramdp-main.py$ / $gramdp-analysis.py$ for GRAM-DP and $DP-main.py$ / $main-DP-analysis.py$ for DP-framework, and provide the path to your dataset along with other required parameters. Make sure that your data is pre-processed (does not contain NaN or empty cells).

\printbibliography
\end{document}